\documentclass[10pt,conference]{IEEEtran}
\IEEEoverridecommandlockouts
\usepackage{cite}
\usepackage{amsmath,amssymb,amsfonts}
\usepackage{amsthm}
\usepackage{amsbsy}
\usepackage{bm}
\usepackage{booktabs}
\usepackage{tabularx}
\usepackage{multicol}
\usepackage{multirow}
\usepackage{caption}
\usepackage{subcaption}
\usepackage{graphicx}
\usepackage{textcomp}
\usepackage{color}
\usepackage{xcolor}
\usepackage{algorithm}
\usepackage{algpseudocode}
\algblock{Output}{EndOutput}
\algnotext{EndOutput}
\algnewcommand{\Initialize}[1]{%
	\State \textbf{Initialize:}
	\Statex \hspace*{\algorithmicindent}\parbox[t]{.8\linewidth}{\raggedright #1}
}
\usepackage{dblfloatfix}
\usepackage[outdir=./]{epstopdf}
\usepackage[left=15.9mm, right=15.9mm, top=19mm,bottom=25.4mm]{geometry}
\usepackage{enumitem}
\newlist{steps}{enumerate}{1}
\setlist[steps, 1]{label = Step \arabic*:}
\allowdisplaybreaks
\def\BibTeX{{\rm B\kern-.05em{\sc i\kern-.025em b}\kern-.08em
\usepackage{hyperref}
    T\kern-.1667em\lower.7ex\hbox{E}\kern-.125emX}}

\DeclareMathOperator{\Tr}{Tr}
\newtheorem{proposition}{Proposition} 
\theoremstyle{remark}

    \usepackage[nodisplayskipstretch]{setspace}
\newcommand{\esp}{\mathbb{E}}

\begin{document}

\title{{AoA-Based Physical-Layer Authentication for CF-mMIMO Systems}

\author{

\IEEEauthorblockN{Isabella W. G. da Silva$^*$, Zahra Mobini$^+$, Hien Quoc Ngo$^*$, and Michail Matthaiou$^*$}

\IEEEauthorblockA{$^*$Centre for Wireless Innovation (CWI), Queen's University Belfast, U.K.}
\IEEEauthorblockA{$^+$Department of Electrical and Electronic Engineering, The University of Manchester, U.K.}

\IEEEauthorblockA{E-mails: \{iwgdasilva01, hien.ngo, m.matthaiou\}@qub.ac.uk; zahra.mobini@manchester.ac.uk}

\vspace{-0.45cm}
}

\thanks{This work was supported by the U.K. Engineering and Physical Sciences
Research Council (EPSRC) grant (EP/X04047X/2) for TITAN Telecoms Hub. The work of I. W. G. da Silva and  H.~Q.~Ngo was supported by a research grant from the Department for the Economy Northern Ireland under the US-Ireland R\&D Partnership Programme. The work of M. Matthaiou was supported by the European Research Council (ERC) under the European Union’s Horizon 2020 research and innovation programme (grant agreement No. 101001331).}
\vspace{-0.5em}}

\maketitle

\begin{abstract}
Cell-free massive multiple-input multiple-output (CF-mMIMO) systems are vulnerable to pilot spoofing attacks, in which an illegitimate user equipment (UE) attempts to reuse pilot sequences to masquerade as a genuine UE. To address this vulnerability, we propose an angle-of-arrival (AoA)-based physical-layer authentication (PLA) framework that exploits the distributed architecture of CF-mMIMO to construct a spatial fingerprint for each UE. To mitigate the pilot contamination introduced by the attacker, the access points (APs) are partitioned into UE- and attacker-dominant subsets, from which the AoAs of the legitimate UE and the attacker are estimated using maximum-likelihood (ML) and low-complexity forward-backward spatial smoothing  Multiple Signal Classification (FBSS-MUSIC) estimators. The authentication problem is formulated as a binary hypothesis test, and closed-form expressions for the probability of detection (PD) and probability of missed detection (PMD) are derived. Numerical results demonstrate that the proposed framework effectively mitigates impersonation attacks and that its authentication performance improves with the number of distributed APs.
\end{abstract}
\begin{IEEEkeywords}
Angle-of-arrival, cell-free massive MIMO, physical-layer authentication (PLA), probability of detection, pilot contamination.
\end{IEEEkeywords}
\vspace{-4mm}
\section{Introduction}\label{sec:Introduction}
\vspace{-0.5em}
CF-mMIMO has emerged as one of the key enablers for next-generation wireless networks. By deploying a large number of spatially distributed APs, CF-mMIMO enables multiple users to be served coherently over the same time-frequency resources, providing ubiquitous coverage and high spectral efficiency~\cite{7827017,10522673}. However, these performance gains rely on accurate channel state information (CSI), which is typically acquired through uplink pilot transmissions~\cite{Mohammadi:PROC.2024}. Consequently, CF-mMIMO systems are particularly vulnerable to pilot spoofing attacks, in which an eavesdropper exploits the publicly known and standardized pilot sequences to impersonate a legitimate UE, thereby contaminating the channel estimation process~\cite{6151778}.

Physical-layer security (PLS) techniques have been proposed to mitigate pilot spoofing attacks~\cite{10666028,10684479}. For instance, the authors in~\cite{10666028} developed a deep neural network to detect pilot spoofing attacks and identify the legitimate UE whose information is being leaked. In~\cite{10684479}, a signal-to-interference-plus-noise ratio (SINR)-maximizing precoding scheme combined with an AP selection strategy based on large-scale fading coefficients was proposed to improve the security performance against spoofing attacks. More recently, PLA has attracted considerable research interest as a lightweight mechanism for verifying transmitter identities using distinctive physical-layer characteristics~\cite{9279294,11003929}. Unlike conventional cryptographic approaches, PLA authenticates the source of a received signal by exploiting distinctive physical-layer characteristics, making it particularly attractive for defending against spoofing attacks~\cite{11432211}. 

A variety of physical-layer attributes have been explored for transmitter authentication, including carrier frequency offset (CFO)~\cite{10930486}, temporal variations of the received SINR~\cite{10834505}, and AoA~\cite{11447296}. Among these attributes, AoA is particularly well suited to CF-mMIMO systems because it is directly related to the spatial position of the transmitter relative to each AP. In conventional co-located systems, spatially separated transmitters may exhibit similar AoAs at the receiver, limiting the discriminative capability of AoA-based authentication. By contrast, CF-mMIMO greatly benefits from its macro-diversity characteristics, in which the authentication feature is observed across spatially distributed receivers, so that the identity of the legitimate UE is encoded by all of the per-AP AoAs. Consequently, an eavesdropper attempting to impersonate a legitimate UE must simultaneously reproduce the AoAs observed at all APs, making successful impersonation attacks increasingly difficult as the number of distributed APs grows.

Despite these promising characteristics, to the best of the authors' knowledge, AoA-based PLA has not yet been investigated for CF-mMIMO systems. Motivated by this research gap, this paper proposes an AoA-based PLA framework for authenticating legitimate UEs against spoofing attacks. The main contributions are summarized as follows:
\begin{itemize}
\item We propose an AoA-based PLA framework for CF-mMIMO systems to authenticate a legitimate UE against pilot spoofing attacks. The APs are first partitioned into UE-dominant and attacker-dominant subsets, after which the AoAs of the legitimate UEs and the eavesdropper are estimated using ML and FBSS-MUSIC estimators.
\item We formulate the authentication decision as a binary hypothesis test and derive closed-form expressions for the PD and PMD of the impersonated UE.
\item Numerical results demonstrate the robustness of the proposed framework against impersonation attacks, showing that the authentication performance greatly benefits from the micro- and macro-diversity gains of CF-mMIMO.
\end{itemize}

\textit{Notation. } Bold upper-case letters denote matrices whereas bold lower-case letters denote vectors; $(\cdot)^T$ and $(\cdot)^H$ stand for the matrix transpose and Hermitian transpose, respectively; $\otimes$ stands for the Kronecker product; $\mathrm{Tr}(\cdot)$ is the trace operator; $\|\cdot\|$ and $|\cdot|$ are the Euclidean-norm and the absolute value operator; $\esp\{\cdot\}$ is the expectation operator, and $\mathbb{R}\{\cdot\}$ represents a real-value extractor. Moreover, $\mathcal{CN}(\mu, \mathbf{C})$ represents a complex Gaussian distribution with mean $\mu$ and covariance matrix $\mathbf{C}$, $\gamma(\cdot,\cdot)$ is the lower incomplete gamma function and $\Gamma(\cdot, \cdot)$ is the upper incomplete gamma function. Finally,  $\chi_k^2$ and $\chi^{`2}_{k}(\mu)$ are, respectively, the central and non-central chi-squared distributions with $k$ degrees of freedom and mean $\mu$.
\vspace{-0.5em}
\section{Authentication Model}
\vspace{-0.5em}
In this section, we describe the system model, followed by uplink training and authentication feature estimation.
\vspace{-0.5em}
\subsection{System Model}
We consider a CF-mMIMO-based PLA scheme with $K$ legitimate UEs and $M$ APs, in the presence of an illegitimate UE, denoted as Eve. The legitimate UEs and Eve are each equipped with a single antenna, whereas the APs are equipped with isotropic half-wavelength-spaced antennas deployed in uniform planar arrays (UPAs) with  $N= N_{h}N_{v}$ antennas, where $N_{h}$  and $N_{v}$ represent the number of antennas in the horizontal and vertical dimensions, respectively. All APs are connected to a central processing unit (CPU); the CPU performs the authentication decisions in a centralized manner.

\begin{figure}[t]
\centering
   \begin{subfigure}[t]{0.4\textwidth}
     \includegraphics[width=\textwidth]{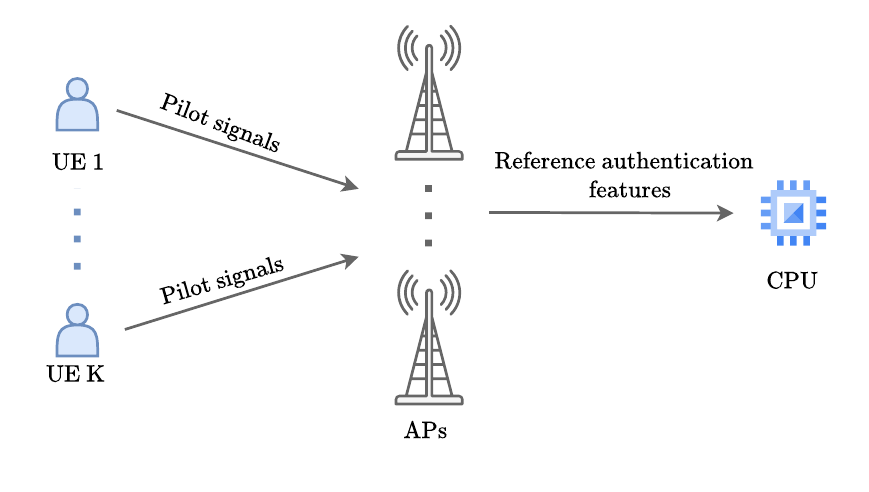}
     \caption{Enrollment phase.}
     \label{fig:enrollment}
   \end{subfigure}
   \begin{subfigure}[t]{0.4\textwidth}
    \includegraphics[width=\textwidth]{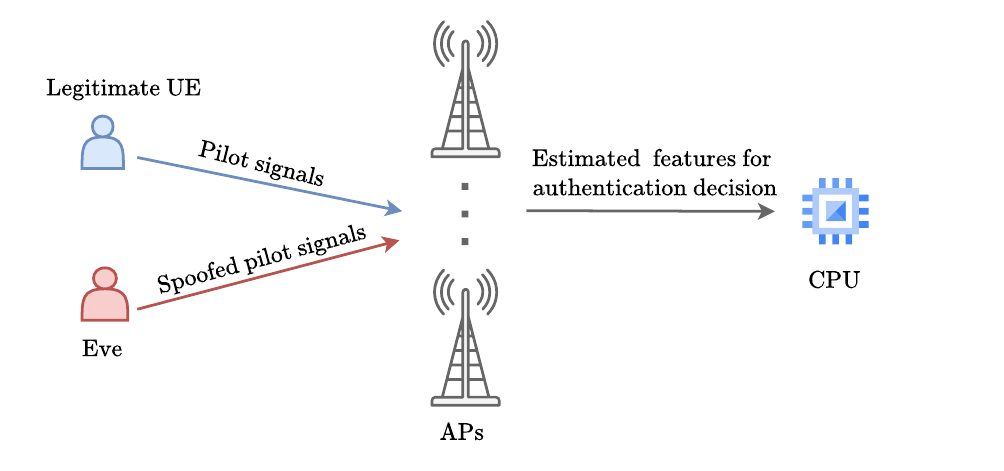}
     \caption{Authentication feature estimation phase.}
     \label{fig:authentication}
   \end{subfigure}
  \caption{System model of the proposed PLA scheme for CF-mMIMO systems.}
  \vspace{0.5em}
\label{fig:phases_PLA}
\end{figure}

The considered PLA scheme is divided into two phases: an enrollment phase and an authentication feature estimation phase, as illustrated in Fig.~\ref{fig:phases_PLA}\subref{fig:enrollment} and Fig.~\ref{fig:phases_PLA}\subref{fig:authentication}, respectively. During the enrollment phase, the $K$ legitimate UEs transmit pilot signals to the APs, which extract the corresponding authentication features and forward them to the CPU, where they are stored as reference identities. The transmissions during the enrollment phase are assumed to be authenticated using upper-layer security mechanisms. During the subsequent authentication phase, the legitimate UEs again transmit pilot signals, which are employed by the APs for uplink training, while Eve simultaneously transmits spoofed pilots in an attempt to impersonate a legitimate UE. The APs estimate the authentication features from the received pilots and forward them to the CPU, which compares them against the stored references to decide whether each received signal originates from a legitimate UE or from an unauthenticated source.

\vspace{-0.5em}
\subsection{Uplink Training}\label{sec:uptraining}
All legitimate UEs requesting uplink access are assumed to transmit pilot sequences to the $M$ APs, with $\bm{\varphi}_k \in \mathbb{C}^{\tau_{\text{p}}\times 1}$ denoting the pilot sequence transmitted by the $k$th UE to the APs, where $\tau_{\text{p}}$ is the pilot sequence length.  All pilot sequences are considered pairwise orthonormal, i.e., $\bm{\varphi}_k^H\bm{\varphi}_{k'}=0$ if $k\neq k'$, and $\|\bm{\varphi}_k\|^2=1$, $\forall k \in \mathcal{K}$, where $\mathcal{K} \triangleq \{1,\dots,K\}$ is the set of legitimate UEs. Thus, it is required that $\tau_{\text{p}}\geq K$.

Simultaneously, Eve aims to launch an impersonation attack or disrupt the authentication process of one of the legitimate UEs by transmitting a signal, such that the resulting received signal at the APs is statistically indistinguishable from that of the impersonated UE. Precisely, we consider the worst-case scenario, where Eve transmits a pilot sequence, denoted by $\bm{\varphi}_{\mathrm{E}}$, that matches the pilot sequence sent by one of the $K$ UEs. This can be accomplished because the network operates under known protocols, with the pilot sequences publicly known and specified in the standard~\cite{6151778}.
Without loss of generality, we assume that Eve transmits the same pilot sequence as UE $1$, i.e., $\bm{\varphi}_{\mathrm{E}} = \bm{\varphi}_{1}$. Hence, the $N\times \tau_p$ received pilot matrix at the $m$th AP is given by
\begin{align}\label{eq:yuplink}
    \mathbf{Y}_{m} = \sum_{k\in \mathcal{K}} \sqrt{\tau_{\text{p}}\rho_{k}}\mathbf{g}_{mk}\bm{\varphi}_{k}^H + \sqrt{\tau_{\text{p}}\rho_{\mathrm{E}}}\mathbf{g}_{m\mathrm{E}}\bm{\varphi}_{1}^H + \bm{\Omega}_{m},
\end{align}
where $\rho_{k}$ and $\rho_{\mathrm{E}}$ are the normalized transmit powers for the $k$th user and for Eve, respectively. Also, $\bm{\Omega}_{m} \in \mathbb{C}^{N\times \tau_{\text{p}}}$ is the noise matrix at the $m$th AP, assumed to have independent and identically distributed (i.i.d.) $\mathcal{CN}(0,1)$ entries. In addition, $\mathbf{g}_{mk}$ and $\mathbf{g}_{m\mathrm{E}} \in \mathbb{C}^{N\times 1}$ are the channel vectors from the $k$th user to the $m$th AP, and from Eve to the $m$th AP, assumed to follow a local scattering and spatially correlated model, and are modeled respectively, as~\cite{bjornson2017massive}
\begin{align}\label{eq:gmodel}
    \mathbf{g}_{mk} &= \mathbf{h}_{mk}(\beta_{mk}\mathbf{R}_{mk})^{1/2},\\
    \mathbf{g}_{m\mathrm{E}} &= \mathbf{h}_{m\mathrm{E}}(\beta_{m\mathrm{E}}\mathbf{R}_{m\mathrm{E}})^{1/2},
\end{align}
where $\mathbf{h}_{mi}, i \in \{k, \mathrm{E}\}$ is the small-scale fading vector, with i.i.d. $\mathcal{CN}(0,1)$ entries, $\beta_{mi}, i \in \{k, \mathrm{E}\}$ is the large-scale fading coefficient, and $\mathbf{R}_{mi}, i \in \{k, \mathrm{E}\}$ is the spatial correlation matrix, given by
\begin{align}
    \mathbf{R}_{m i}=\mathbf{R}_{m i}^h(\phi_{m i}, \theta_{m i}) \otimes \mathbf{R}_{m i}^v(\theta_{m i}),
\end{align}
where $\phi_{mi}$ and $\theta_{mi}$ are the azimuth and elevation AoAs at the $m$th AP UPA, respectively. Finally,  $\mathbf{R}_{m i}^h(\phi_{m i}, \theta_{m i})$ and  $\mathbf{R}_{m i}^v(\theta_{m i})$ are written as
\begin{align}
[\mathbf{R}_{m i}^h]_{n_h, n_h^{\prime}}&= e^{j \pi\Delta_h \sin({\phi}_{m i}) \sin ({\theta}_{m i})}\nonumber\\
&\quad\times e^{-\frac{\pi^2 \sigma_\phi^2 \cos ^2({\phi}_{m i}) \sin ^2({\theta}_{m i})}{2}\Delta_h^2}, \\
[\mathbf{R}_{m i}^v]_{n_v, n_v^{\prime}}&= e^{j \pi\Delta_v \cos ({\theta}_{m i})}e^{-\frac{\pi^2 \sigma_\theta^2 \sin ^2({\theta}_{m i})}{2}\Delta_v^2},
\end{align}
where $n_h = \{1,\dots,N_{h}\}$ and $n_v = \{1,\dots,N_{v}\}$, $\Delta_h \triangleq (n_h-n_h^{\prime})$, $\Delta_v \triangleq (n_v-n_v^{\prime})$, and $\sigma_\phi$ and $\sigma_\theta$ are the azimuth and elevation angular standard deviations, respectively. The $m$th AP then projects the received pilot signal in~\eqref{eq:yuplink} onto $\bm{\varphi}_k$ for the estimation of the channel vector from user $k$ to the $m$th AP, which is given by
\begin{align}\label{eq:tildey}
     \Tilde{\mathbf{y}}_{mk} &= \begin{cases}
         \sqrt{\tau_{\text{p}}\rho_{k}}\mathbf{g}_{mk}+ \Tilde{\bm{\omega}}_{mk}, &k \neq 1,\\
         \sqrt{\tau_{\text{p}}\rho_{1}}\mathbf{g}_{m1}+ \sqrt{\tau_{\text{p}}\rho_{\mathrm{E}}}\mathbf{g}_{m\mathrm{E}}+\Tilde{\bm{\omega}}_{m1}, &k = 1,
     \end{cases}
\end{align}
where $\Tilde{\bm{\omega}}_{mk} \triangleq \bm{\Omega}_{m}\bm{\varphi}_{k}$, $\forall k \in \mathcal{K}$. Based on~\eqref{eq:tildey},  a linear minimum mean-squared error (MMSE) approach can be employed to obtain estimates of the channel vectors $\mathbf{g}_{mk}$ as
\begin{align}
    \hat{\mathbf{g}}_{mk} &= \esp\{\mathbf{g}_{mk}\Tilde{\mathbf{y}}_{mk}\}\left(\esp\left\{\Tilde{\mathbf{y}}_{mk}\Tilde{\mathbf{y}}_{mk}^H\right\}\right)^{-1}\Tilde{\mathbf{y}}_{mk}\nonumber\\
    &=\sqrt{\tau_{\text{p}}\rho_{k}}\beta_{mk}\mathbf{R}_{mk}\bm{\Psi}_{mk}^{-1}\Tilde{\mathbf{y}}_{mk},
\end{align}
where $\bm{\Psi}_{mk}$ is given by
\begin{align}
     \bm{\Psi}_{mk} =\begin{cases}
        \tau_{\text{p}}\rho_{k}\beta_{mk}\mathbf{R}_{mk}\! +\! \mathbf{I}_{N}, & k \neq 1,\\
        \tau_{\text{p}}\rho_{1}\beta_{m1}\mathbf{R}_{m1}\! +\! \tau_{\text{p}}\rho_{\mathrm{E}}\beta_{m\mathrm{E}}\mathbf{R}_{m\mathrm{E}}\! +\! \mathbf{I}_{N}, & k = 1.
    \end{cases} 
\end{align}
Thus, $\hat{\mathbf{g}}_{mk} \sim \mathcal{CN}\left(\mathbf{0},\bm{\Gamma}_{mk}\right)$, with
\begin{align}
    \bm{\Gamma}_{mk} =\begin{cases}
        \tau_{\text{p}}\rho_{k}\beta_{mk}^2\mathbf{R}_{mk}\bm{\Psi}_{mk}^{-1}\mathbf{R}_{mk} , & k \neq 1,\\
        \tau_{\text{p}}\rho_{1}\beta_{m1}^2\mathbf{R}_{m1}\bm{\Psi}_{m1}^{-1}\mathbf{R}_{m1}, & k = 1.
    \end{cases} 
\end{align}
Accordingly, the channel estimation error $\bm{\epsilon}_{mk}$ $\sim$ $ \mathcal{CN}(\mathbf{0},\beta_{mk}\mathbf{R}_{mk} - \bm{\Gamma}_{mk})$, and is independent of $\hat{\mathbf{g}}_{mk}$.
\vspace{-0.5em}
\subsection{Authentication Feature Estimation}
The APs located in the vicinity of Eve may receive a pilot observation containing significant contributions from both UE 1 and Eve, which can degrade the reliable estimation of the legitimate UE's authentication feature. As a result, these APs cannot provide reliable estimates of the authentication feature to the CPU. Accordingly, let $\mathcal{M}$ denote the set of APs, with $|\mathcal{M}| = M$. We propose an AP selection scheme that partitions $\mathcal{M}$ into two complementary subsets, with each contributing to a distinct feature estimation that is later used for the authentication decision of the legitimate UEs at the CPU. Specifically, the $c$th AP is assigned to a subset $\mathcal{C}_k$\footnote{The CPU has no prior knowledge of which legitimate UE is being impersonated by Eve. Thus, the AP selection and subsequent feature estimation should be evaluated for all $k \in \mathcal{K}$.}, if for each $k$th UE in the network, the ratio $\Tilde{\beta}_{ck}/\beta_{ck}$ exceeds a predetermined threshold,  where $\tilde{\beta}_{ck}$ is the effective large-scale fading coefficient of the $k$th UE at the $c$th AP, obtained by measuring the average received power of the projected signal 
$\tilde{\mathbf{y}}_{ck}$ in \eqref{eq:tildey}. Specifically, since $\tilde{\mathbf{y}}_{ck}$ has i.i.d. entries whose variance scales with the large-scale fading, $\tilde\beta_{ck}$ can be estimated as
$\tilde\beta_{ck} = (\frac{1}{N}\|\tilde{\mathbf{y}}_{ck}\|^2 - 1)/\tau_p \rho_k$. Otherwise, the $c$th AP is assigned to $\mathcal{C}_k^{\mathrm{c}}\triangleq \mathcal{M} \setminus \mathcal{C}_k$.
Then, the AoA from the legitimate UEs is estimated by the APs in $\mathcal{C}_k$. Simultaneously, the APs in $\mathcal{C}_k^c$ also estimate the AoA from Eve. Accordingly, let $\bar{\phi}_{i}^{\min} \leq \bar{\phi}_{i} \leq \bar{\phi}_{i}^{\max}$
and $\bar{\theta}_{i}^{\min} \leq \bar{\theta}_{i} \leq
\bar{\theta}_{i}^{\max}$, $i \in \{c,c'\}$, $\forall c \in \mathcal{C}_k$ and $\forall c' \in \mathcal{C}_k^c$, delimit a search grid at the $i$th AP for the
azimuth and elevation AoAs, respectively. The estimates of the AoAs are derived considering the ML and the MUSIC approaches, as outlined next.

\subsubsection{ML-based AoA estimation}
To obtain the ML estimates of the AoA from the $k$th UE to the $c$th AP in $\mathcal{C}_k$, we consider that based on the AP selection criterion, the contribution of Eve to the projected pilot signal in \eqref{eq:tildey} is negligible in comparison to that of the $k$th UE, i.e., $\tau_p\rho_{\mathrm{E}}\beta_{c\mathrm{E}}\mathbf{R}_{c\mathrm{E}} \approx \mathbf{0}, \forall c \in \mathcal{C}_k$. Under this assumption, $\Tilde{\mathbf{y}}_{ck} \sim \mathcal{CN}(\mathbf{0}, \bm{\Psi}_{ck}), \forall k \in \mathcal{K}$, with probability density function (PDF) for a given search grid point $(\bar{\phi}_{ck}, \bar{\theta}_{ck})$ written as
\begin{align}\label{eq:ptildey}
    p_{\Tilde{\mathbf{y}}_{ck}} =\frac{\exp\{-\Tilde{\mathbf{y}}_{ck}^H\bm{\Psi}_{ck}^{-1}(\bar{\phi}_{ck},\bar{\theta}_{ck})\Tilde{\mathbf{y}}_{ck}\}}{\pi^{N}\det(\bm{\Psi}_{ck}(\bar{\phi}_{ck},\bar{\theta}_{ck}))}.
\end{align}
Based on \eqref{eq:ptildey}, the log-likelihood function is given by
\begin{align}\label{eq:Lmk}
    \mathcal{L}_{ck}(\bar{\phi}_{c},\bar{\theta}_{c}) &= \log p_{\Tilde{\mathbf{y}}_{ck}}\nonumber\\
    &= - \tilde{\mathbf{y}}_{ck}^H \bm{\Psi}_{ck}^{-1}(\bar{\phi}_c, \bar{\theta}_c) \tilde{\mathbf{y}}_{ck}
- N\log \pi\nonumber\\
&\quad- \log \det (\bm{\Psi}_{ck}(\bar{\phi}_c, \bar{\theta}_c)).
\end{align}
Thus, the ML estimate of the AoA from the $k$th UE to the $c$th AP, denoted by $(\hat{\phi}_{ck}^{\mathrm{ML}}, \hat{\theta}_{ck}^{\mathrm{ML}})$ is obtained by selecting the search grid point that maximizes \eqref{eq:Lmk}, as
\begin{align}
(\hat{\phi}_{ck}^{\mathrm{ML}}, \hat{\theta}_{ck}^{\mathrm{ML}}) &= \arg\max_{\bar{\phi}_c, \bar{\theta}_c}
\mathcal{L}_{ck}(\bar{\phi}_c, \bar{\theta}_c), \quad c \in \mathcal{C}_k.
\end{align}
On the other hand, for all $c'$ in $\mathcal{C}_k^c$, we have that $\tilde{\beta}_{c'k}/\beta_{c'k}$ is larger than the threshold, which implies that the contribution of Eve dominates in \eqref{eq:tildey}. Thus, it is feasible to assume that $\tau_p\rho_{k}\beta_{c'k}\mathbf{R}_{c'k} \approx \mathbf{0}$. Hence, $\Tilde{\mathbf{y}}_{c'k} \sim \mathcal{CN}(\mathbf{0}, \bm{\Psi}_{c'\mathrm{E}}), \forall k \in \mathcal{K}$. Then, the $ c'$-th AP can construct an estimate of the covariance of Eve for a point of the search grid by considering that
\begin{align}
    \rho_{\mathrm{E}}\beta_{c'\mathrm{E}} \approx \frac{\tilde{\beta}_{c'k}}{\tau_{\mathrm{p}}} - \frac{\rho_{k}\beta_{c'k}}{\tau_{\mathrm{p}}}.
\end{align}
Finally, $(\hat{\phi}_{c'\mathrm{E}}^{\mathrm{ML}},\hat{\theta}_{c'\mathrm{E}}^{\mathrm{ML}})$ can be attained as 
\begin{align}
(\hat{\phi}_{c'\mathrm{E}}^{\mathrm{ML}}, \hat{\theta}_{c'\mathrm{E}}^{\mathrm{ML}}) &= \arg\max_{\bar{\phi}_{c'}, \bar{\theta}_{c'}}
\mathcal{L}_{c'\mathrm{E}}(\bar{\phi}_{c'}, \bar{\theta}_{c'}), \quad c' \in \mathcal{C}_k^{\mathrm{c}},
\label{eq:mle_eve}
\end{align}
where
\begin{align}
\mathcal{L}_{c'\mathrm{E}}(\bar{\phi}_{c'}, \bar{\theta}_{c'}) &=
- \tilde{\mathbf{y}}_{c'k}^H \hat{\bm{\Psi}}_{c'\mathrm{E}}^{-1}(\bar{\phi}_{c'}, \bar{\theta}_{c'}) \tilde{\mathbf{y}}_{c'k}
- N \log \pi\nonumber\\
&\quad- \log \det (\hat{\bm{\Psi}}_{c'\mathrm{E}}(\bar{\phi}_{c'}, \bar{\theta}_{c'})).
\label{eq:loglik_eve}
\end{align}

\textbf{Computational Complexity:} Note that $(\hat{\phi}_{ck}^{\mathrm{ML}}, \hat{\theta}_{ck}^{\mathrm{ML}})$ and $(\hat{\phi}_{c'\mathrm{E}}^{\mathrm{ML}},\hat{\theta}_{c'\mathrm{E}}^{\mathrm{ML}})$ are obtained via an exhaustive search over a two-dimensional angular grid. For each grid point, it is necessary to invert an $N\times N$ matrix, with a complexity cost of $\mathcal{O}(N^3)$. Accordingly, the total computational complexity of $(\hat{\phi}_{ck}^{\mathrm{ML}}, \hat{\theta}_{ck}^{\mathrm{ML}})$ and of $(\hat{\phi}_{c'\mathrm{E}}^{\mathrm{ML}},\hat{\theta}_{c'\mathrm{E}}^{\mathrm{ML}})$ is given by $\mathcal{O}(N^3\Delta_{\bar{\phi}_i}\Delta_{\bar{\theta}_i})$, $i \in \{c,c'\}$,  where $\Delta_{\bar{\phi}_i}=(\bar{\phi}_i^{\mathrm{max}}-\bar{\phi}_i^{\mathrm{min}})$ and $\Delta_{\bar{\theta}_i}=(\bar{\theta}_i^{\mathrm{max}}-\bar{\theta}_i^{\mathrm{min}})$.
\subsubsection{MUSIC-based AoA estimation}
Given the computational cost of the ML-based AoA estimation, we consider an alternative AoA estimation approach based on the MUSIC algorithm. Specifically, given that the covariance of the projected pilot signal in \eqref{eq:tildey} is not full rank, the standard MUSIC algorithm cannot be directly employed to obtain the estimates of $(\phi_{ck},\theta_{ck})$ and of $(\phi_{c'\mathrm{E}}, \theta_{c'\mathrm{E}})$. Under this consideration, we first perform a FBSS on \eqref{eq:tildey}, as in~\cite{17496} and~\cite{10413280}, to partition the UPA of all APs in $\mathcal{M}$ into overlapping sub-arrays of size $N' = N_{h}'N_{v}$, where $N_{h}'\leq N_{h}$ and $N_{v}'\leq N_{v}$, with the total number of sub-arrays set as $L = L_hL_v$. Thus, $L_h$ and $L_v$ are subjected to $L_h = N_{h} - N_{h}' + 1$ and $L_v = N_{v} - N_{v}' + 1$. Accordingly, we define a spatial smoother as $\mathcal{S}(\mathbf{x}) \triangleq  [\mathbf{J}_{1}\mathbf{x},\dots,\mathbf{J}_{l}\mathbf{x},\dots,\mathbf{J}_{L}\mathbf{x}]$, in which $\mathbf{J}_{l} = \mathbf{J}_{l_h}\otimes \mathbf{J}_{l_v}$, $\forall l_h \in \{1, L_h\}$ and $\forall l_v \in \{1, L_v\}$. Specifically, $\mathbf{J}_{l_h}$ and $\mathbf{J}_{l_v}$ are selection matrices of dimensions $N_{h}'\times N_{h}$ and $N_{v}'\times N_{v}$, respectively, and are written as 
\begin{align}
    \mathbf{J}_{l_h} &= [\mathbf{0}_{N_{h}'\times (l_h-1)}, \mathbf{I}_{N_{h}'},\mathbf{0}_{N_{h}'\times (N_{h}- N_{h}'-l_h+1)}],\\
     \mathbf{J}_{l_v} &= [\mathbf{0}_{N_{v}'\times (l_v-1)}, \mathbf{I}_{N_{v}'},\mathbf{0}_{N_{v}'\times (N_{v}- N_{v}'-l_v+1)}].
\end{align}
Next, let $\bm{\Pi} \triangleq \bm{\Pi}_h \otimes \bm{\Pi}_v$ define a $N' \times N'$ exchange matrix. The spatially smoothed forward and backward signals are, respectively, given by $\mathbf{Y}_{ck}^f \triangleq \mathcal{S}(\tilde{\mathbf{y}}_{ck})$ and $\mathbf{Y}_{ck}^b \triangleq \bm{\Pi}\mathbf{Y}_{ck}^{f*}$. Following, we construct 
$\mathbf{Y}_{ck}^{fb} = \frac{1}{\sqrt{2L}
}[\mathbf{Y}_{ck}^f, \mathbf{Y}_{ck}^b] \in \mathbb{C}^{N'\times 2L}$, $\forall c \in \mathcal{C}_k$, and $\mathbf{Y}_{c'k}^{fb} = \frac{1}{\sqrt{2L}
}[\mathbf{Y}_{c'k}^f, \mathbf{Y}_{c'k}^b] \in \mathbb{C}^{N'\times 2L}$, $\forall c' \in \mathcal{C}^c_k$, which are, respectively, employed for the estimation of $(\phi_{ck},\theta_{ck})$ and of $(\phi_{c'\mathrm{E}}, \theta_{c'\mathrm{E}})$. Accordingly, for every $c \in \mathcal{C}_k$ and $c' \in \mathcal{C}_k^c$, we perform a singular value decomposition (SVD) as follows
\begin{align}
    \mathbf{Y}_{ck}^{fb} &=  \mathbf{E}_{ck}\Lambda_{ck}\mathbf{T}_{ck}^H,\\
    \mathbf{Y}_{c'k}^{fb} &=  \mathbf{E}_{c'\mathrm{E}}\Lambda_{c'\mathrm{E}}\mathbf{T}_{c'\mathrm{E}}^H,
\end{align}
where $\Lambda_{ck}$ and $\Lambda_{c'\mathrm{E}}$ are non-zero singular value diagonal matrices of size $r \times r$, with $r \leq \min\{N', 2L\}$ being the rank of $\mathbf{Y}_{ck}^{fb}$ and of $\mathbf{Y}_{c'k}^{fb}$. Also, $\mathbf{E}_{i} \in \mathcal{C}^{N'\times r}$ and $\mathbf{T}_{i} \in \mathcal{C}^{2L\times r}$, $i \in \{ck, c'\mathrm{E}\}$ are semi-unitary matrices, such that $\mathbf{E}_{i}^H\mathbf{E}_{i} = \mathbf{T}_{i}^H\mathbf{T}_{i} = \mathbf{I}$. Precisely, knowing that $\mathbf{E}_{ck}$ and $\mathbf{E}_{c'\mathrm{E}}$, respectively, span the column space of $\mathbf{Y}_{ck}^{fb}$ and  $\mathbf{Y}_{c'k}^{fb}$, which is correspondent to the antenna domain responses, we partition $\mathbf{E}_{ck}$ and $\mathbf{E}_{c'\mathrm{E}}$ as 
\begin{align}
    \mathbf{E}_{ck} &\triangleq [\mathbf{E}_{ck}^S, \mathbf{E}_{ck}^N],\\
    \mathbf{E}_{c'\mathrm{E}} &\triangleq [\mathbf{E}_{c'\mathrm{E}}^S, \mathbf{E}_{c'\mathrm{E}}^N],
\end{align}
where $\mathbf{E}_{ck}^S$ and $\mathbf{E}_{c'\mathrm{E}}^S \in \mathbb{C}^{N'\times D}$ are subspaces corresponding to the $D$ dominant singular values, and $\mathbf{E}_{ck}^N$ and $\mathbf{E}_{c'\mathrm{E}}^N \in \mathbb{C}^{N'\times (N'- D)}$ are noise subspaces. Consequently, the FBSS-MUSIC pseudo-spectra over the search grid can be constructed as 
\begin{align}
    P_{ck}(\bar{\phi}_{c},\bar{\theta}_{c}) &= \frac{1}{\mathbf{a}^H(\bar{\phi}_{c},\bar{\theta}_{c})\mathbf{E}_{ck}^N(\mathbf{E}_{ck}^N)^H\mathbf{a}(\bar{\phi}_{c},\bar{\theta}_{c})},\\
    P_{c'\mathrm{E}}(\bar{\phi}_{c'},\bar{\theta}_{c'}) &= \frac{1}{\mathbf{a}^H(\bar{\phi}_{c'},\bar{\theta}_{c'})\mathbf{E}_{c'\mathrm{E}}^N(\mathbf{E}_{c'\mathrm{E}}^N)^H\mathbf{a}(\bar{\phi}_{c'},\bar{\theta}_{c'})},
\end{align}
where $\mathbf{a}(\bar{\phi}_c, \bar{\theta}_c) = \mathbf{a}_h(\bar{\phi}_c,\bar{\theta}_c)\otimes \mathbf{a}_v(\bar{\phi}_c,\bar{\theta}_c) \in \mathbb{C}^{N'\times 1}$ is the sub-array steering vector, in which,
\begin{subequations}\label{eq:arbarc}
\begin{align}
    \left[\mathbf{a}(\bar{\phi}_c, \bar{\theta}_c)\right]_{n_h'} &=\exp(j\pi(n_h'-1)\sin(\bar{\phi}_c)\sin(\bar{\theta}_c)),\\
    \left[\mathbf{a}(\bar{\theta}_c)\right]_{n_v'} &=\exp(j\pi(n_v'-1)\cos(\bar{\theta}_c)).
\end{align}
\end{subequations}
Finally,
\begin{align}
(\hat{\phi}_{ck}^{\mathrm{SS}}, \hat{\theta}_{ck}^{\mathrm{SS}}) &= \arg\max_{\bar{\phi}_c, \bar{\theta}_c}
P_{ck}(\bar{\phi}_{c},\bar{\theta}_{c}), \quad c \in \mathcal{C}_k,\\
(\hat{\phi}_{c'\mathrm{E}}^{\mathrm{SS}}, \hat{\theta}_{c'\mathrm{E}}^{\mathrm{SS}}) &= \arg\max_{\bar{\phi}_{c'}, \bar{\theta}_{c'}}
P_{c'\mathrm{E}}(\bar{\phi}_{c'},\bar{\theta}_{c'}), \quad c' \in \mathcal{C}^c_k,
\end{align}

\textbf{Computational Complexity:} The computational cost is dominated by the construction of the FBSS matrices $\mathbf{Y}_{ck}^{fb}$ and $\mathbf{Y}_{c'k}^{fb}$, and by the computation of the pseudo-spectrums over the search grid, with a total complexity of $\mathcal{O}(LNN' + N'^2\Delta_{\bar{\phi}_c}\Delta_{\bar{\theta}_c})$ for obtaining $(\hat{\phi}_{ck}^{\mathrm{SS}}, \hat{\theta}_{ck}^{\mathrm{SS}})$ and $\mathcal{O}(LNN' + N'^2\Delta_{\bar{\phi}_{c'}}\Delta_{\bar{\theta}_{c'}})$ for $(\hat{\phi}_{c'\mathrm{E}}^{\mathrm{SS}},\hat{\theta}_{c'\mathrm{E}}^{\mathrm{SS}})$.
\vspace{-0.8em}
\section{Authentication Phase}
\vspace{-0.5em}
After obtaining the AoA pair estimate for the $k$th UE, $(\hat{\phi}_{ck},\hat{\theta}_{ck})$, and of Eve, $(\hat{\phi}_{c'\mathrm{E}},\hat{\theta}_{c'\mathrm{E}})$, the APs in $\mathcal{C}_k$ and in $\mathcal{C}_k^c$ forward their corresponding estimates to the CPU, which performs a binary hypothesis test as
\begin{align}\label{eq:autdes}
    \left\{\begin{array}{l}
\mathcal{H}_0^{\mathrm{aut}}: \text{the AoAs originate from the $k$th UE}, \\
\mathcal{H}_1^{\mathrm{aut}}: \text{the AoAs do not originate from the $k$th UE}.
\end{array}\right. 
\end{align}
Then, the CPU makes the authentication decision via a threshold test between $\mathcal{H}_0^{\mathrm{aut}}$ and $\mathcal{H}_1^{\mathrm{aut}}$ for each user $k$, that is, 
\begin{align}\label{eq:threshold}
    \lambda_k \stackrel{\mathcal{H}_1^{\mathrm{aut}}}{\underset{\mathcal{H}_0^{\mathrm{aut}}}{\gtrless}} \eta_k,
\end{align}
where $\eta_k$ is the decision threshold of the $k$th user, and $\lambda_k$ is the test statistic, computed as~\cite{kay1998fundamentals}
\begin{align}\label{eq:test}
    \lambda_k =  \sum_{c \in \mathcal{C}_k} \mathbf{r}_{ck}^{T}\mathbf{C}_{ck}^{-1}\mathbf{r}_{ck},
\end{align}
with $\mathbf{r}_{ck}$ being the angular residual at the $c$th AP, given by
\begin{align}
    \mathbf{r}_{ck} = \bigg[\begin{array}{c}
\hat{\phi}_{c k}-\phi_{c k}^* \\
\hat{\theta}_{c k}-\theta_{c k}^*
\end{array}\bigg],
\end{align}
 where $(\phi_{c k}^*,\theta_{c k}^*)$ are the reference AoAs of the $k$th UE at the $c$th AP, obtained during the enrollment phase, assumed to accurately represent the true AoAs of the $k$th UE. In addition, $\mathbf{C}_{ck}$ is the covariance of the estimate evaluated at $(\hat{\phi}_{ck}, \hat{\theta}_{ck})$, computed in~\eqref{eq:weighting}. 
Based on the threshold test in~\eqref{eq:threshold}, if the CPU accepts $\mathcal{H}_0^{\mathrm{aut}}$ when $\mathcal{H}_0^{\mathrm{aut}}$ is true, UE $k$ has been correctly detected. On the other hand, if $\mathcal{H}_0^{\mathrm{aut}}$ is accepted when $\mathcal{H}_1^{\mathrm{aut}}$ is true, a missed detection has occurred. Thus, the expressions for the PD and PMD are derived in the following proposition. 
\begin{proposition}
    Based on the authentication decision in \eqref{eq:threshold}, the PD and PMD of the $k$th UE are, respectively, given by
    \begin{align}
        \mathrm{P}_{\mathrm{D},k} &= \frac{\gamma(|\mathcal{C}_k|, \eta_k/2)}{\Gamma(|\mathcal{C}_k|)},\label{eq:PDK}\\
        \mathrm{P}_{\mathrm{MD},k} &= 1 - Q_{|\mathcal{C}_k|}(\sqrt{\mu_k},\sqrt{\eta_k})\label{eq:PMD},
    \end{align}
    where $Q_{|\mathcal{C}_k|}(\cdot,\cdot)$ is the generalized Marcum-Q function of order $|\mathcal{C}_k|$, and $\mu_k$ is given by
    \begin{align}
        \mu_k = \sum_{c \in \mathcal{C}_k} \mathbf{r}_{c\mathrm{E}}^T \mathbf{C}_{ck}^{-1}\mathbf{r}_{c\mathrm{E}},
    \end{align}
    where $\mathbf{r}_{c\mathrm{E}}$ is the angular offset from the estimate of the AoAs of Eve to the reference AoA of the $k$th UE at the $c$th AP in $\mathcal{C}_k$, that is, defined as
    \begin{align}
        \mathbf{r}_{c\mathrm{E}} = \bigg[\begin{array}{c}
        \hat{\phi}_{c\mathrm{E}}-\phi_{c k}^* \\
        \hat{\theta}_{c\mathrm{E}}-\theta_{c k}^*
\end{array}\bigg].
    \end{align}
\end{proposition}
\begin{proof}
    The proof is provided in Appendix A.
\end{proof}
\vspace{-1.3em}
\section{Numerical Results}\label{sec:Results}
\vspace{-0.5em}
In this section, we assess the performance of the proposed CF-mMIMO AoA-based PLA scheme. To this end, we consider a scenario in which the APs and legitimate UEs are randomly distributed within a $1\times 1$ km$^2$ area with wrapped-around edges to avoid boundary effects. The channel bandwidth is $B=20$ MHz and the pilot length is $\tau_{\text{p}} = 30$. The maximum transmit power of each legitimate UE and Eve is set to $200$ mW. The corresponding normalized transmit powers, $\rho_{k}$ and $\rho_{\mathrm{E}}$, can be obtained by dividing the transmit power by the noise power, given by
\begin{align}
    \text{noise power} = B \times k_B \times T_0 \times \text{noise figure }  (\mathrm{W}),
\end{align}
where $k_B = 1,381 \times 10^{-23}$ (Joule per Kelvin) is the Boltzmann constant, $T_0 = 290$ (Kelvin) is the noise temperature, and the noise figure is set as 7.94. The large-scale fading coefficients $\beta_{mk}$ and $\beta_{m\text{E}}$ are computed as
\begin{align}
    \beta_{\mathrm{l}} = 10^{\frac{-\mathrm{PL}_{\mathrm{l}}+\sigma_{\mathrm{sh}}z_{\mathrm{l}}}{10}}, ~\mathrm{l} \in \{mk, m\text{E}\},
\end{align}
where $10^{\frac{\sigma_{\mathrm{sh}}z_{\mathrm{l}}}{10}}$ represents the shadow fading with the standard deviation $\sigma_{\mathrm{sh}} = 7.8$ dB, and $z_{\mathrm{l}} \sim \mathcal{N}(0, 1)$, and $\mathrm{PL}_{\mathrm{l}}$ represents the path loss, computed as~\cite{3GPP}
\begin{align}
    \mathrm{PL}_{\mathrm{l}}= 32.4 + 20\log_{10}(f) + 30\log_{10}(d_{3\mathrm{D}}),
\end{align}
where $f = 1.9$ GHz is the carrier frequency, and $d_{3\mathrm{D}} = \sqrt{(d_l)^2 + (h_{\mathrm{AP}} - h_{\mathrm{u}})^2}$ is the three-dimensional distance between APs and users measured in meters, with $d_l$ denoting the horizontal distance between the nodes. Specifically, to compute $d_{3\mathrm{D}}$, we assume that the antenna height at the APs is set to  $h_{\mathrm{AP}} = 15$ m, and the antenna height of both the legitimate UEs and Eve is set to  $h_{\mathrm{u}} = 1.65$ m, $u \in \{k, \mathrm{E}\}$. The decision threshold $\eta_k$ is computed from~\eqref{eq:PMD} for a fixed missed detection rate, $p_{\mathrm{MD},k}$, as
\begin{align}\label{eq:etakopt}
    \eta_k^* = \big(Q_{|\mathcal{C}_k|}^{-1}(\sqrt{\mu_k},1-p_{\mathrm{MD},k})\big)^2,
\end{align}
where, unless otherwise stated, $p_{\mathrm{MD},k}$ is set as $10^{-2}$. Finally, $\sigma_\phi$ and $\sigma_\theta$ are set to $0.2$ rad.

\begin{figure}[t]
    \centering
    \includegraphics[width=0.35\textwidth]{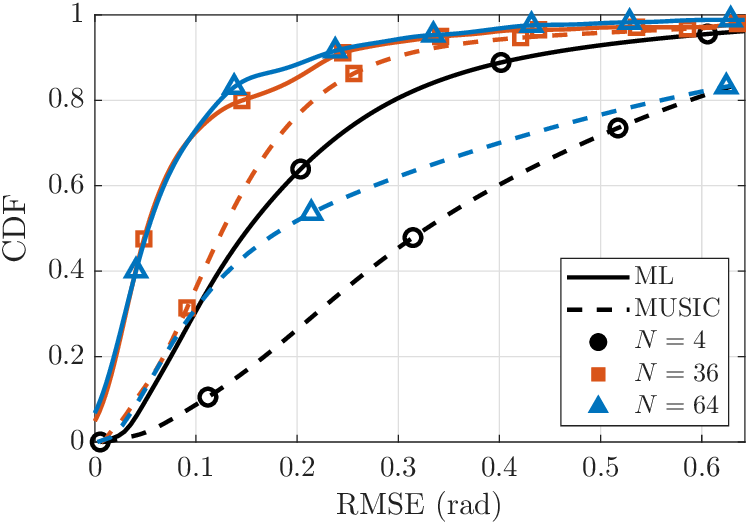}
    \caption{CDF of the  RMSE of the AoA of Eve.}
    \label{fig:RMSE}
\end{figure}

To evaluate the accuracy of the ML and FBSS-MUSIC estimators, and the proposed AP selection scheme, Fig.~\ref{fig:RMSE} illustrates the cumulative distribution function (CDF) of the root-mean-square-error (RMSE) of the AoA of Eve at one AP $c' \in \mathcal{C}_k^c$ for different values of $N$.  Note that, since the test statistic in~\eqref{eq:test} is normalized by $\mathbf{C}_{ck}$, a lower per-AP RMSE directly impacts on the PD of the impersonated UE. Accordingly, as expected, note from Fig.~\ref{fig:RMSE} that the ML estimator outperforms the FBSS MUSIC estimator for all evaluated values of $N$. Specifically, we note that the RMSE of the ML estimator improves substantially from $N=4$ to $N=36$. On the other hand, the FBSS MUSIC estimator performance approaches that of the ML for $N=36$, and the performance degrades as $N$ increases from $N=36$ to $N=64$, which is due to the fact that the MUSIC algorithm is based on the assumption of single point sources. However, as the antenna array size increases, the impact of the angular spread from the local scattering channel model becomes non-negligible, and the single point-source steering vector assumed by MUSIC no longer matches the received signal.



\begin{figure}[t]
    \centering
    \includegraphics[width=0.35\textwidth]{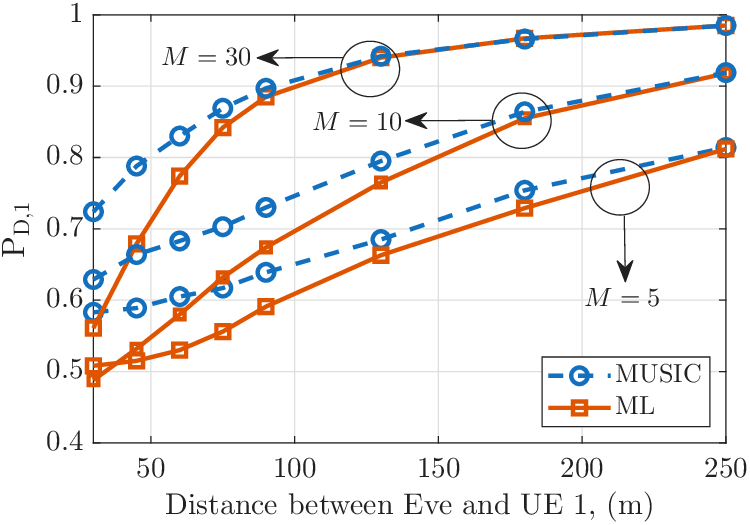}
    \caption{$\mathrm{P}_{\mathrm{D},1}$ versus the distance between Eve and UE $1$ ($\mathrm{m}$), with $N = 36$.}
    \label{fig:PD_Distance}
\end{figure}

Finally, to evaluate the impact of the macro-diversity gain provided by CF-mMIMO systems on the proposed PLA framework,  Fig.~\ref{fig:PD_Distance} illustrates $\mathrm{P}_{\mathrm{D},1}$ versus the distance between Eve and the impersonated UE for different values of $M$. We observe that the framework significantly benefits from an increased number of APs, with a performance gain of approximately $50$\% as $M$ increases from $5$ to $30$. Furthermore, as previously mentioned, in CF-mMIMO systems, Eve needs to be located as close to the $k$th UE as possible for the attack to be effective, since the authentication feature is observed across spatially distributed APs. Nonetheless, we note that for the case of $M=30$, the proposed framework is capable of achieving a satisfactory $\mathrm{P}_{\mathrm{D},k}$ even when the distance between Eve and the impersonated UE is small. On this matter, we observe that further performance gains may be possible by  employing active security mechanisms against Eve, such as the transmission of jamming signals or combining multiple authentication features, which we identify as an interesting direction for future
work.   

\section{Conclusion}\label{sec:Conclusions}
We proposed a CF-mMIMO AoA-based PLA scheme framework to authenticate a  legitimate UE against pilot spoofing attacks. In our system, the APs are partitioned into UE-dominant and Eve-dominant subsets, and the AoA of both the legitimate UEs and the attacker is estimated using ML and FBSS-MUSIC estimators. We formulated a hypothesis testing problem and derived closed-form expressions for the PD and PMD for each UE. The numerical results showed that the proposed CF-mMIMO AoA-based PLA framework is robust against spoofing attacks and accurate in detecting the AoA of Eve. They also showed that the detection performance further improves as the number of deployed APs and the number of antennas per AP increase, confirming that the degrees of freedom offered by CF-mMIMO can be effectively exploited for PLA applications. Potential future work includes extending the framework to scenarios with a multi-antenna Eve, as well as to scenarios with multiple attackers and UE mobility.

\vspace{-0.4em}
\section*{Appendix A \\Proof of Proposition~1}
\vspace{-0.4em}
The PD and PMD for the $k$th UE are given by
\begin{align}
    \mathrm{P}_{\mathrm{D},k} = \Pr(\lambda_k< \eta_k | \mathcal{H}_0^{\mathrm{aut}}),\\
     \mathrm{P}_{\mathrm{MD},k} = \Pr(\lambda_k< \eta_k | \mathcal{H}_1^{\mathrm{aut}}),
\end{align}
respectively. Given that, for the MLE and FBSS MUSIC estimation, the AoA estimates follow a Gaussian distribution asymptotically~\cite{kay1998fundamentals}, we have that, under $\mathcal{H}_0^{\mathrm{aut}}$, $(\hat{\phi}_{ck},\hat{\theta}_{ck})$ converges in probability to $(\phi_{ck}^*,\theta_{ck}^*)$. Consequently, $\mathbf{r}_{ck}|\mathcal{H}_0^{\mathrm{aut}} \sim \mathcal{N}(\mathbf{0},\mathbf{C}_{ck}(\phi_{ck}^*,\theta_{ck}^*))$. Specifically, 
 \begin{align}
 \mathbf{C}_{ck} = \begin{cases}
 \mathbf{C}_{ck}^{\mathrm{ML}}(\phi_{ck}^*,\theta_{ck}^*) & \text{(ML)} ,\\
 \mathbf{C}_{ck}^{\mathrm{SS}}(\phi_{ck}^*,\theta_{ck}^*) & \text{(FBSS MUSIC)},
 \end{cases}
 \label{eq:weighting}
 \end{align}
 where $\mathbf{C}_{ck}^{\mathrm{ML}}(\phi_{ck}^*,\theta_{ck}^*)$ is computed as the inverse of the Fisher information matrix, $\mathbf{I}_{ck}(\phi_{ck}^*,\theta_{ck}^*)$, which is attained as
 \begin{align}
     \mathbf{I}_{ck}(\phi_{ck}^*,\theta_{ck}^*)=\left[\begin{array}{cc}
 \Tr(AA) & \Tr(AB) \\
 \Tr(BA) & \Tr(BB)
 \end{array}\right],
 \end{align}
 where $A = \bm{\Psi}_{ck}^{-1}\frac{\partial \bm{\Psi}_{ck}}{\partial {\phi}_{ck}^*}$ and $B= \bm{\Psi}_{ck}^{-1}\frac{\partial \bm{\Psi}_{ck}}{\partial {\theta}_{ck}^*}$, with $\frac{\partial \bm{\Psi}_{ck}}{\partial {\phi}_{ck}^*}$ and $\frac{\partial \bm{\Psi}_{ck}}{\partial {\theta}_{ck}^*}$ are respectively given by
 \begin{align}
     \frac{\partial \bm{\Psi}_{c k}}{\partial \phi_{ck}^*}&=\tau_{\mathrm{p}} \rho_{k} \beta_{c k}\left(\frac{\partial \mathbf{R}_{c k}^h}{\partial \phi_{ck}^*} \otimes \mathbf{R}_{c k}^v\right),\\
     \frac{\partial \bm{\Psi}_{c k}}{\partial \theta_{ck}^*}&=\tau_{\mathrm{p}} \rho_{k} \beta_{c k}\left(\frac{\partial \mathbf{R}_{c k}^h}{\partial \theta_{ck}^*} \otimes \mathbf{R}_{c k}^v+\mathbf{R}_{c k}^h \otimes \frac{\partial \mathbf{R}_{c k}^v}{\partial \theta_{ck}^*}\right).
 \end{align}
 Finally, $\frac{\partial \mathbf{R}_{c k}^h}{\partial \phi_{ck}^*}$, $\frac{\partial \mathbf{R}_{c k}^h}{\partial \theta_{ck}^*}$ and $\frac{\partial \mathbf{R}_{c k}^v}{\partial \theta_{ck}^*}$ are, respectively, written as
 \begin{align}
     \left[\frac{\partial \mathbf{R}_{c k}^h}{\partial \phi_{ck}^*}\right]_{n_h, n_h^{\prime}}&\!\!\!=\left[\mathbf{R}_{c k}^h\right]_{n_h, n_h^{\prime}}(j \pi \cos(\phi_{ck}^*) \sin(\theta_{ck}^*) \Delta_h\nonumber\\
     &+\pi^2 \sigma_\phi^2 \sin(\phi_{ck}^*) \cos(\phi_{ck}^*) \sin ^2(\theta_{ck}^*) \Delta_h^2),\\
     \left[\frac{\partial \mathbf{R}_{c k}^h}{\partial \theta_{ck}^*}\right]_{n_h, n_h^{\prime}}&\!\!\!=\left[\mathbf{R}_{c k}^h\right]_{n_h, n_h^{\prime}}(j \pi \sin(\phi_{ck}^*) \cos(\theta_{ck}^*) \Delta_h\nonumber\\
     &-\pi^2 \sigma_\phi^2 \cos^2(\phi_{ck}^*) \sin(\theta_{ck}^*) \cos(\theta_{ck}^*) \Delta_h^2),\\
     \left[\frac{\partial \mathbf{R}_{c k}^v}{\partial \theta_{ck}^*}\right]_{n_v, n_v^{\prime}}&=\left[\mathbf{R}_{c k}^v\right]_{n_v, n_v^{\prime}}(-j \pi \sin(\theta_{ck}^*) \Delta_v\nonumber\\
     &\quad-\pi^2 \sigma_\theta^2 \sin(\theta_{ck}^*) \cos(\theta_{ck}^*) \Delta_v^2).
 \end{align}
 Furthermore, $\mathbf{C}_{ck}^{\mathrm{SS}}(\phi_{ck}^*, \theta_{ck}^*)$ is obtained as~\cite{197097}
 \begin{align}
     \mathbf{C}_{ck}^{\mathrm{SS}}(\phi_{ck}^*, \theta_{ck}^*) = \frac{1}{4L(\tau_p\rho_{k}\beta_{ck})^2}\mathbb{R}(\mathbf{D}_{ck}^H\mathbf{D}_{ck})^{-1},
 \end{align}
where
\begin{align}
    \mathbf{D}_{ck} &= \bigg(\mathbf{I}_{N'}-\frac{\mathbf{a}(\phi_{ck}^*, \theta_{ck}^*)\mathbf{a}^H(\phi_{ck}^*, \theta_{ck}^*)}{N^{\prime}}\bigg)\nonumber\\
    &\quad\times\bigg[\frac{\partial \mathbf{a}(\phi_{ck}^*, \theta_{ck}^*)}{\partial \phi_{ck}^*}, \frac{\partial \mathbf{a}(\phi_{ck}^*, \theta_{ck}^*)}{\partial \theta_{ck}^*}\bigg].
\end{align}
Under these considerations, and by rewriting $\lambda_k$ as
\begin{align}
    \lambda_k = \sum_{c\in \mathcal{C}_k}\|\mathbf{C}_{ck}^{-1/2}\mathbf{r}_{ck}\|^2,
\end{align}
 we have that $\lambda_k| \mathcal{H}_0^{\mathrm{aut}} \sim \chi^2_{2|\mathcal{C}_k|}$, since the term $\|\mathbf{C}_{ck}^{-1/2}\mathbf{r}_{ck}\|^2$ is independent for all $c$ in $\mathcal{C}_k$. Therefore, the $\mathrm{P}_{\mathrm{D},k}$ is given by the CDF of $\lambda_k| \mathcal{H}_0^{\mathrm{aut}}$, as in \eqref{eq:PDK}.

 On the other hand, to derive $\mathrm{P}_{\mathrm{MD},k}$, we start by computing the AoA estimates between Eve and the  $c$th AP in $\mathcal{C}_k$, $(\hat{\phi}_{c\mathrm{E}},\hat{\theta}_{c\mathrm{E}})$, from the estimates between Eve and the APs in $\mathcal{C}_k^c$. For that, the CPU estimates the position of Eve via a least-squares problem as
 \begin{align}
     \hat{\mathbf{p}}_{\mathrm{E}} &= \arg\min_{\mathbf{p}}
\sum_{c' \in \mathcal{C}_k^{\mathrm{c}}}
\bigg(\bigg[\begin{array}{c}
\hat{\phi}_{c'\mathrm{E}}-\phi_{c'p} \\
\hat{\theta}_{c'\mathrm{E}}-\theta_{c'p}
\end{array}\bigg]^T
\mathbf{C}_{c'\mathrm{E}}^{-1}(\hat{\phi}_{c'\mathrm{E}},\hat{\theta}_{c'\mathrm{E}})\nonumber\\
&\quad\times\bigg[\begin{array}{c}
\hat{\phi}_{c'\mathrm{E}}-\phi_{c'p} \\
\hat{\theta}_{c'\mathrm{E}}-\theta_{c'p}
\end{array}\bigg]\bigg),
\label{eq:triangulation}
 \end{align}
where $(\phi_{c'p},\theta_{c'p})$ is the AoA from AP $c'$ from position $\mathbf{p}$, and $\mathbf{C}_{c'\mathrm{E}}^{-1}(\hat{\phi}_{c'\mathrm{E}},\hat{\theta}_{c'\mathrm{E}})$ is the estimated inverse covariance of Eve at the $c'$th AP, computed as in~\eqref{eq:weighting}. Then, given $\hat{\mathbf{p}}_{\mathrm{E}}$, the AoA of Eve at the $m$th AP in $\mathcal{C}_k$ is set as $(\hat{\phi}_{c\mathrm{E}},\hat{\theta}_{c\mathrm{E}}) \triangleq ({\phi}_{c}(\hat{\mathbf{p}}_{\mathrm{E}}),{\theta}_{c}(\hat{\mathbf{p}}_{\mathrm{E}}))$.  Thus, under $\mathcal{H}_1^{\mathrm{aut}}$, we can consider that $(\hat{\phi}_{ck},\hat{\theta}_{ck})$ converges to $(\hat{\phi}_{c\mathrm{E}},\hat{\theta}_{c\mathrm{E}})$. Thus, $\mathbf{r}_{ck}|\mathcal{H}_1^{\mathrm{aut}} \sim \mathcal{N}(\mathbf{r}_{c\mathrm{E}},\mathbf{C}_{ck}(\phi_{ck}^*,\theta_{ck}^*))$, and $\lambda_k| \mathcal{H}_1^{\mathrm{aut}} \sim \chi^{`2}_{2|\mathcal{C}_k|}(\mu_k)$. Hence, $\mathrm{P}_{\mathrm{MD},k}$ is given by the CDF of the $\chi^{`2}_{2|\mathcal{C}_k|}(\mu_k)$ distribution, as in~\eqref{eq:PMD}.
\vspace{-0.5em}
\bibliographystyle{IEEEtran}  
\bibliography{references}

\end{document}